\documentclass[10pt, conference]{IEEEtran}

\IEEEoverridecommandlockouts
\usepackage{amsmath,amssymb,amsfonts}

\usepackage{tcolorbox}
\usepackage{multirow}
\usepackage{url}
\usepackage{enumitem}
\usepackage{diagbox}
\usepackage{caption}

\def\BibTeX{{\rm B\kern-.05em{\sc i\kern-.025em b}\kern-.08em
    T\kern-.1667em\lower.7ex\hbox{E}\kern-.125emX}}
\begin{document}






\title{Automated Generating Natural Language Requirements based on Domain Ontology}



\author{Ziyan Zhao, Li Zhang, Xiaoyun Gao, Xiaoli Lian, Heyang LV, Lin Shi}

\maketitle

\begin{abstract}
    Software requirements specification is undoubtedly critical for the whole software life-cycle. Nowadays, writing software requirements specifications primarily depends on human work. Although massive studies have been proposed to fasten the process via proposing advanced elicitation and analysis techniques, it is still a time-consuming and error-prone task that needs to take domain knowledge and business information into consideration. In this paper, we propose an approach, named \emph{ReqGen}, which can provide recommendations by automatically generating natural language requirements specifications based on certain given keywords. Specifically, \emph{ReqGen} consists of three critical steps. First, keywords-oriented knowledge is selected from domain ontology and is injected to the basic Unified pre-trained Language Model (UniLM) for domain fine-tuning. Second, a copy mechanism is integrated to ensure the occurrence of keywords in the generated statements. Finally, a requirement syntax constrained decoding is designed to close the semantic and syntax distance between the candidate and reference specifications. Experiments on two public datasets from different groups and domains show that \emph{ReqGen} outperforms six popular natural language generation approaches with respect to the hard constraint of keywords(phrases) inclusion, BLEU, ROUGE and syntax compliance. We believe that \emph{ReqGen} can promote the efficiency and intelligence of specifying software requirements.
    

\end{abstract}






\section{Introduction}
\label{sec:intro}

There is no doubt about the importance of software requirements to the whole software life cycle \cite{Terzakis2013TheIO}\cite{REManaging}\cite{REMicrosoft}. As the vital product of requirement analysis, software requirement specifications act as the essential bridge between the requirement analysis stage and the following development and test. 
Nowadays, writing software requirements specifications primarily relies on human work, and this work is complex and time-consuming due to the following factors.

\begin{itemize}[leftmargin=1em]
    \item Enough domain knowledge is required for stating the right content as well as selecting appropriate words and expressions. But domain analysis typically requires non-trivial human effort \cite{DBLP:conf/re/LianRCZFS16}.
    
    \item Writing the specifications word-by-word is time-consuming, let alone that many expressions are repeated, especially in similar or related requirements. Writing or locating, copying, and then pasting the repeated content is a waste of human effort.
    
    \item Generally accepted requirement syntax, such as EARS \cite{5328509}, is suggested for writing well-formed specifications. Learning and carefully applying non-business-related knowledge is also a burden on requirement analysts.
    
\end{itemize}

Intuitively, it would be quite helpful to automatically recommend requirements specifications as long as the analysts can provide some simple related information almost effortlessly. To the best of our knowledge, most of the automated requirements specifications generation work focus on transforming software engineering models (e.g., i* framework \cite{DBLP:journals/re/MaidenMJG05} \cite{DBLP:conf/coopis/YuBDM95}, KAOS \cite{DBLP:conf/sigsoft/LetierL02} \cite{ DBLP:journals/re/LandtsheerLL04} \cite{van2004goal}, UML models\cite{DBLP:journals/tse/LamsweerdeW98} \cite{DBLP:journals/re/MezianeAA08} \cite{DBLP:conf/re/Berenbach03a}) or other semi-structured inputs (e.g., security goals in specific syntactic patterns \cite{10.1007/s00766-017-0279-5}) into natural language requirement specifications based on pre-defined rules, which are usually brittle and restrict the usability of these approaches. What's more, constructing expressive and precise models is another complex work too.

\begin{figure}[htbp]
	\centering
	\includegraphics[trim = {7cm 9cm 10cm 4cm}, clip, width=0.4\textwidth]{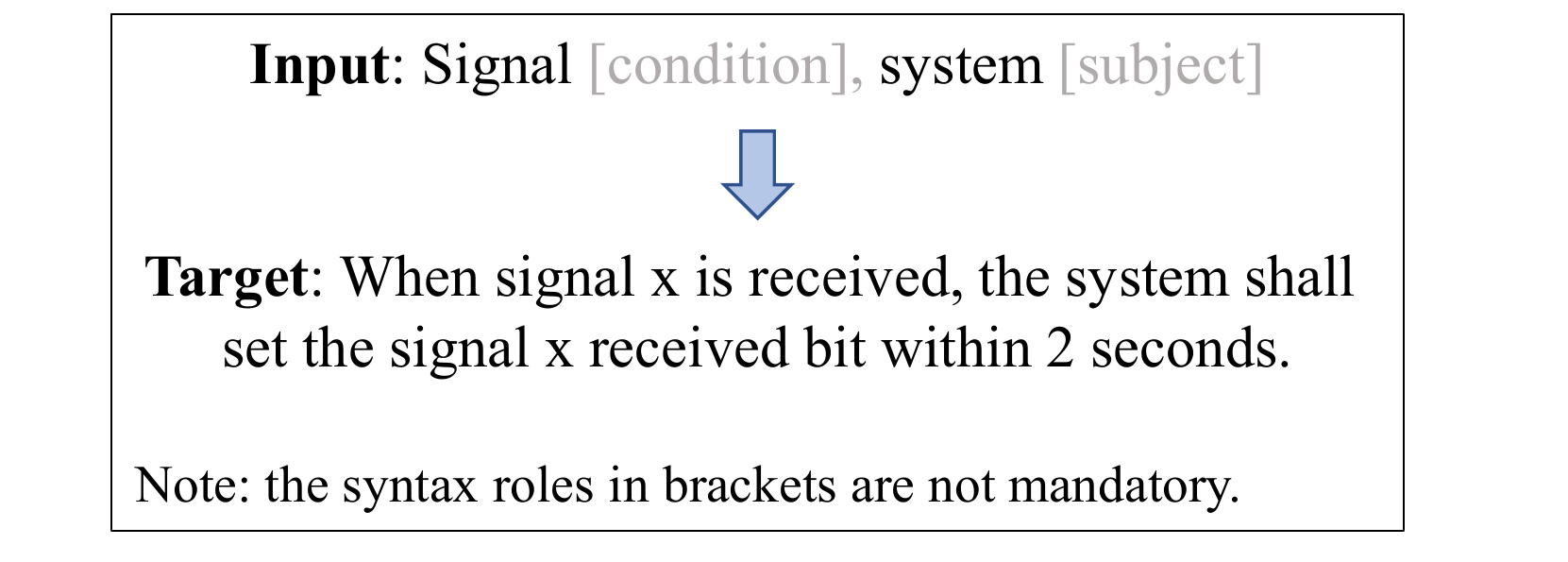} 
    \caption{One example showing the target of this work.}\label{fig:in-out-example}
\end{figure}

Besides, lots of researches have been conducted to speed up the requirement analysis process by identifying and analyzing requirements information from, for example, domain documents \cite{DBLP:conf/re/LianRCZFS16} \cite{DBLP:conf/re/LianCZ17} \cite{DBLP:journals/infsof/LianLZ20} \cite{LI2015582}, developers online chats \cite{9283914}, and product descriptions \cite{DBLP:conf/icse/DumitruGHCMCM11} \cite{DBLP:conf/splc/Sree-KumarPC18}. And the primary results of these studies are features \cite{DBLP:conf/icse/DumitruGHCMCM11} \cite{DBLP:conf/splc/Sree-KumarPC18} \cite{DBLP:conf/re/ZhouHLLPF15} \cite{DBLP:conf/re/KhanXLW19} \cite{DBLP:conf/re/AstegherBPS21} \cite{DBLP:conf/re/JohannSBM17} \cite{8054853}, requirement-related sentences \cite{DBLP:conf/re/LianRCZFS16}\cite{DBLP:conf/re/LianCZ17}\cite{DBLP:journals/infsof/LianLZ20} \cite{DBLP:conf/re/SleimiCSSBD19} \cite{FalknerPFSAS19} \cite{shi2020detection} \cite{DBLP:conf/re/DevineKB21} \cite{DBLP:conf/re/HenaoFSFV21} and requirement classifications \cite{DBLP:conf/re/KhanXLW19} \cite{DBLP:conf/re/SainaniAJG20} \cite{DBLP:conf/re/Tizard19}. Although these resources are relevant and helpful for requirements acquisition and generation, they are only \emph{separate pieces of information}. Requirement analysts still need to spend significant efforts to understand them, to integrate them with the project background, and then to specify the final requirements by following at least one requirement syntax\cite{DBLP:journals/infsof/LianLZ20}. 
In this work, we aim to automatically generate requirement statements once the analyst has an intuitive idea and can provide two or more keywords (phrases) of the desired requirements. Optionally, the analyst can also suggest the syntax roles of part or all keywords. 
We hope to recommend the requirement specifications although requirement analysts probably need to revise our generation for the final acceptance. To illustrate our target task, we contrived one example from a requirement instance of ISO/IEC/IEEE 29148:2018(E) \cite{8559686} shown in Fig. \ref{fig:in-out-example}.

In this paper, we propose an approach, named \emph{ReqGen}, to generate requirements specifications from the keywords (phrases) provided by requirement analysts. In particular, three critical designs are proposed upon the basic Unified pre-trained Language Model (UniLM) \cite{DBLP:conf/nips/00040WWLWGZH19}, which is selected as the backbone of our framework because of its promising performance on natural language generation (NLG) tasks\cite{DBLP:conf/emnlp/LinZSZBCR20}.
First, we inject domain knowledge in the format of pseudo-sentences created from domain ontologies into several selected layers of UniLM. 
Second, we use a copy mechanism in the training phase and set the given keywords as a hard constraint to ensure that they occur in the final statements. 
Last but not least, we design a requirement syntax constrained decoding approach to accommodate to the requirement syntax constraints of the generated statements, given that well-formed requirements should follow a certain syntax \cite{DBLP:conf/re/GuoZL21}\cite{5328509}\cite{8559686}. 

To evaluate our approach, we conduct experiments on public datasets consisting of two domains, which were collected from previous work \cite{DBLP:conf/re/ZhaoZL21}. 
The results demonstrate the promising performance of \emph{ReqGen}, especially on the dataset with a larger knowledge scale.
The comparisons with six popular NLG approaches show that \emph{ReqGen} obtains better Bilingual Evaluation Understudy (BLEU), Recall-Oriented Understudy for Gisting Evaluation (ROUGE) and requirement syntax compliance. We also show the effectiveness of the three proposed components in \emph{ReqGen} using an ablation experiment. The major contributions of this work are as follows.


\begin{itemize}[leftmargin=1em]
    \item An approach \emph{ReqGen} of automatically generating software requirement statements based on two or more keywords. 
    
    \item The evaluation of \emph{ReqGen} on two public datasets from different domains with promising results, and suggestions on knowledge injection to pre-trained models.
    
    \item The source code and experimental data publicly available on Github \url{https://github.com/ZacharyZhao55/ReqGen}.
    
\end{itemize}

\noindent \textbf{\emph{Significance}.} Requirements recommendation is a major concern in RE. Comparing with the traditional approaches, we take a step forward by recommending the final requirements specifications rather than only identifying the useful but indirect information.
\section{Background}
\label{sec:background}

This section describes three key techniques: UniLM, attention mechanisms and bidirectional long short-term memory (Bi-LSTM). 




\subsubsection{\textbf{UniLM}}
\label{subsec:bgUnilm}
We use UniLM \cite{DBLP:conf/nips/00040WWLWGZH19} as the backbone of our approach because of its good NLG performance \cite{DBLP:conf/emnlp/LinZSZBCR20}\cite{DBLP:journals/corr/abs-2012-10813}. UniLM is a multi-layer Transformer network with 786-hidden dimensions, and 16 attention heads. Its parameter size is 340M, and the activation function is Gaussian Error Linear Unit (GeLU), the same as those of Bidirectional Encoder Representation from Transformers(BERT) \cite{devlin2018bert}. It is pre-trained using two unidirectional language models (LM), one bidirectional LM, and one seq2seq LM. For our requirement generation task, we configured it using seq2seq LM (also known as the encoder-decoder model). UniLM uses the 12-layer bidirectional Transformer encoder like BERT. Each encoder layer includes a multi-head attention and a feed-forward neural network. During decoding, UniLM uses beam search (beam size is 5 in our implementation) to select the candidate tokens of top-k scores in each step. 

\subsubsection{\textbf{Attention mechanism}}
\label{subsec:bgAtt}
The model with conventional encoders cannot pay attention to information outside the input sequence. Our aim is to make the model can pay more attention to the injected knowledge; in other words, to let the model change the weights of the original hidden state so that higher weights can be assigned to the injected knowledge. 

Many attention methods exist, including the dot product model, scaled dot product model, additive model, and bi-linear model. Moreover, the steps of these methods are almost the same, including (1) calculation of the similarity between \emph{Query} and \emph{Key}, and obtaining the weights; (2) normalization of the weights; and (3) summation of the weights with \emph{Value}.  The scaled dot product is the most-common, fast, and space-efficient attention mechanism \cite{DBLP:conf/nips/VaswaniSPUJGKP17}, and we use it. This model has an input consisting of a \emph{Query} and \emph{Key} with $d_k$ dimensions, and a \emph{Value} with $d_v$ dimensions. 
We first calculate the dot product of the \emph{Query} and \emph{Key} (MatMul), then divide every \emph{Key} by $\sqrt{d_k}$ (Scale). Finally, we feed it to a \emph{softmax} layer to obtain the weights corresponding to the \emph{Value}s. 



\subsubsection{\textbf{Bi-LSTM}}
\label{subsec:bgbilstm}
To better understand the injected knowledge, contextual information should be considered in the model. Before injection, we need to encode the knowledge. In the method proposed in this paper, we use Bi-LSTM \cite{DBLP:conf/icassp/GravesMH13} to encode the knowledge in order to capture its contextual information from the injected knowledge, where the knowledge is embedded by BERT. Bi-LSTM is a variant of Recurrent Neural Network, which combines two standard LSTM \cite{DBLP:journals/neco/HochreiterS97} layers in opposite directions to learn the two-way representation. An LSTM cell has three gates: the input, forget, and output gates. These three gates have different functions, acting as filters that decide which information to keep and which to forget. This increases the accessible information of the Bi-LSTM network and hence the model can better understand the context of the knowledge.

\section{Our approach: ReqGen}
\label{sec:approach}

\begin{figure*}
\centering
 \includegraphics[trim = {0.5cm 4.1cm 0.4cm 1.2cm}, clip, width=\textwidth]{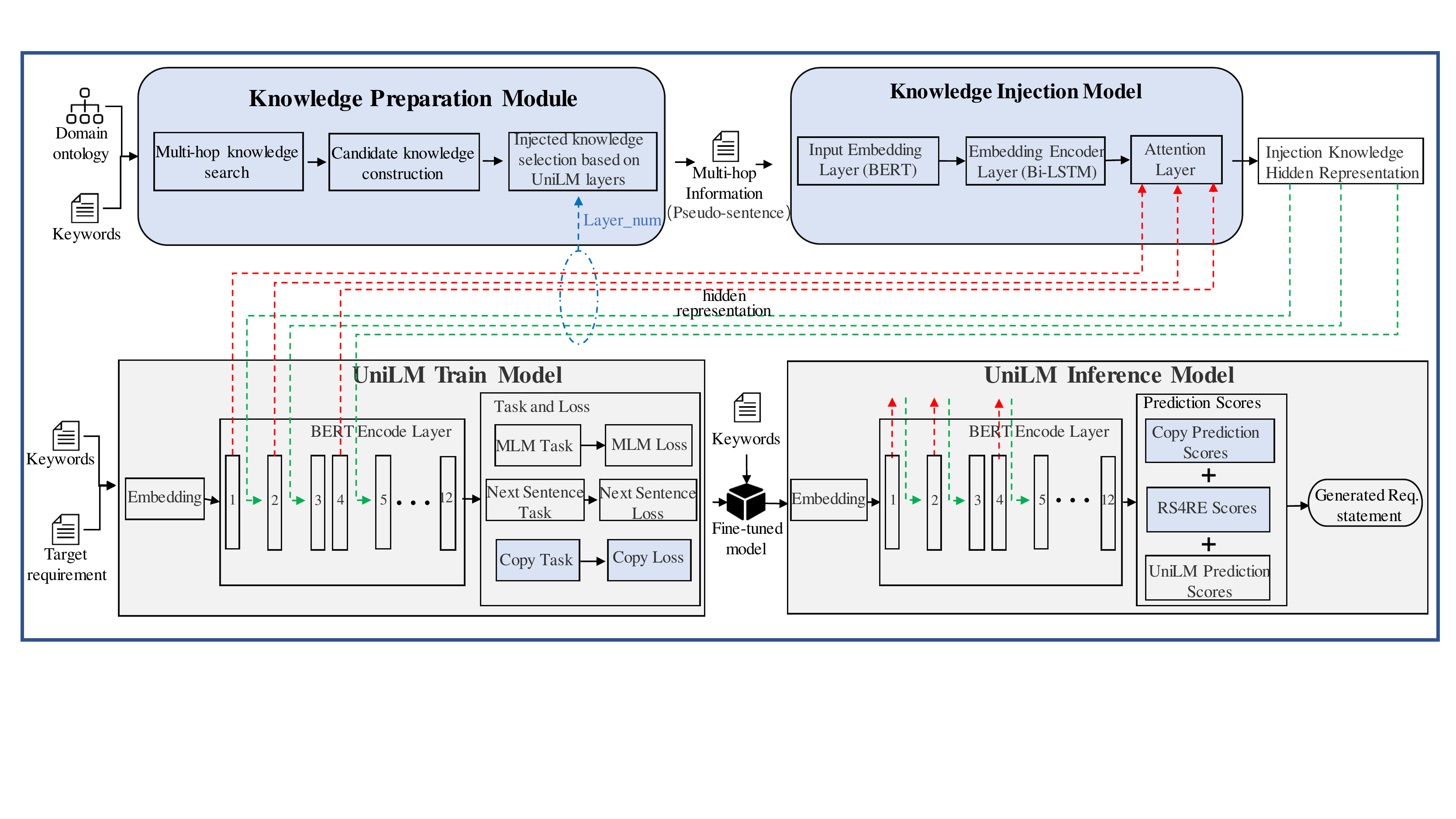} 
\caption{The procedure of our ReqGen}\label{fig:framework}
\end{figure*}

As shown in Fig. \ref{fig:framework}, we select UniLM as the backbone of \emph{ReqGen}, because of its advantages in natural language understanding and NLG \cite{DBLP:conf/nips/00040WWLWGZH19}. On the basis of UniLM, four components are designed and implemented to improve the compliance of the generated statements with domain knowledge and software requirements syntax (as indicated by the light blue rectangles in Fig. \ref{fig:framework}):

\begin{itemize}[leftmargin=1em]
     \item A knowledge preparation module that retrieves keyword-related information from the domain ontology (in terms of triples consisting of entity pairs and their relationships), transforms all related information into pseudo-sentence, and selects the knowledge to be injected into the different layers of UniLM.
    
    \item A knowledge-injection model to inject the knowledge produced by the knowledge preparation module into UniLM. The injected knowledge sequence is encoded by a Bi-LSTM structure using a BERT-based sentence embedding and is injected through the attention mechanism.
    
    \item  A keywords copy mechanism is added in the original UniLM, which enables it to perform a copied-word classification task in the UniLM training. Moreover, a prediction method is added to the UniLM inference model to decide whether the next token is a copied word.
    
    \item A requirement syntax constrained decoding module that includes a semantic-related metric used in the inference model with the original beam search to optimize the final statements towards a specific syntax. 
\end{itemize}

\subsection{Knowledge preparation module}
\label{subsec:knowledgePre}

The purpose of this module is to produce the knowledge to be injected in the backbone UniLM according to the input keywords from the domain ontology. It consists of three steps: multi-hop knowledge search, pseudo-sentence construction, and knowledge selection for the UniLM layers.

\subsubsection{Multi-hop knowledge search}
\label{subsubsec: multi-hopSearching}

This step aims to obtain the keyword-related concepts and their relationships from a domain ontology. We would like to increase the probability of the co-occurrence of these concepts and the input keywords by managing the attention in the knowledge injection model.

We obtain knowledge from the domain ontology, which is a collection of all of the relevant concepts and their relationships in a single domain. It is represented as a graph structure composed of triples $\langle \emph{Entity$_{1}$, Relation, Entity$_{2}$} \rangle$, built based on OWL ontology language rules \cite{bechhofer2004owl}.

To obtain the context information of the keywords, we first perform a multi-hop graph search to acquire as much useful information as possible (five hops in this work), starting from the keywords, and collecting all types of entities in the retrieved paths.

Because all of the input keywords are regarded as start nodes, there must be repeated retrieved paths. We believe the concepts in the repeated paths are important; that is, the more often they are retrieved, the more important they are. Following this principle, we filter some concepts according to the retrieval times to avoid noise. 


\subsubsection{Pseudo-sentence construction} 
We convert the extracted multi-hops $\langle \emph{Entity$_{1}$, Relation, Entity$_{2}$} \rangle$  into pseudo-sentences for the following injection task. In the OWL ontology language, entity types include \emph{Classes}, \emph{Object Properties}, and \emph{Named Individuals}. And there are two types of relations between these entities: the \emph{subclass relation} and \emph{constraint relation}. We set pseudo-sentence generation rules for these two relation types as follows.

(1) The \emph{subclass} type includes \emph{subClassOf} and \emph{subPropertyOf} relations.
\begin{itemize}[leftmargin=1em]
\item \emph{subClassOf} or \emph{hasSuperClasses}: For the triple of \emph{$\langle a, subClassOf, b \rangle$}, we create a sentence of ``a is subclass of b''. And we convert the triple of \emph{$\langle \mathtt{a, hasSuperClasses, b} \rangle$} into ``a has super class b.''
\item \emph{subPropertyOf}: For the triple of \emph{$\langle \mathtt{a, subPropertyOf, b} \rangle$}, we create a sentence like ``a is subproperty of b.''.
\end{itemize}

(2) The \emph{Constraint} type includes \emph{has domain} and \emph{has range} relations.
\begin{itemize}[leftmargin=1em]
\item \emph{has domain}: Specifies the domain of a property \emph{P}, indicating that any resource with a given property is an instance of the domain class (e.g., \emph{$\langle \mathtt{teaching, has domain, teacher} \rangle$}).

\item \emph{has range}: Specifies the range of a property \emph{P}, indicating that the value of a property is an instance of the range class (e.g., \emph{$\langle \mathtt{teaching, has range, lesson} \rangle$}).
\end{itemize}

Generally, the relationships of \emph{has domain} and \emph{has range} are paired. If the property has a \emph{has domain} triple, there must be a corresponding \emph{has range} triple simultaneously. Hence, we produce sentences for the relationship pairs. Taking the triples of \emph{$\langle \mathtt{teaching, has domain, teacher} \rangle$} and \emph{$\langle \mathtt{teaching, has range, lesson} \rangle$ } as examples, we can create the pseudo-sentence ``Teacher is teaching lesson."

Note that the grammar of the pseudo-sentences generated automatically may be wrong. However, it can provide an important context for the concepts in sentences that is required by Bi-LSTM encoder. 


\subsubsection{Knowledge selection towards the UniLM layers} 

{In the human learning process, when learning new stuff, we typically learn relatively broad knowledge first and then pay more attention to the essential parts, which is a repeated and gradual process. Inspired by this, we inject different keyword-related knowledge three times to the UniLM encoder and the injected knowledge is more refined every time, rather than the common one-time knowledge injection into the pre-trained model \cite{DBLP:journals/corr/abs-2012-10813}.}


Following the traditional one-time knowledge injection in the first layer of the pre-trained BERT model \cite{DBLP:journals/corr/abs-2012-10813}, we observed that the injected knowledge cannot be reflected in the decoder of BERT. We believe the primary reason is the strong fitting ability of BERT on training data, i.e., too small weights of the external knowledge injected once than those of training data. 

Therefore, we design a knowledge-injection mechanism by increasing the number of injections and selecting different knowledge for different layers of UniLM. In this work, we inject the knowledge into the $1^{st}, 2^{nd}$, and $4^{th}$ layer of the UniLM encoder, as shown in Fig.\ref{fig:framework}. Moreover, we inject more valuable information into the higher layers. In particular, we select all 5-hop keyword-related information for the first layer, 2-hop information for the second layer, and only 1-hop information for the fourth layer. This gradual refinement of the injected knowledge should enhance the ability of UniLM to absorb externally injected information. The evaluation and analysis of these three layers selection (i.e., 1,2,4) are shown in Section \ref{sec:exptimesofmultiinjection}.

\subsection{Knowledge-injection model}
\label{subsec:KnowledgeInjection}

As Fig. \ref{fig:framework} shows, the knowledge-injection model is composed of three layers. The multi-hop information in the format of pseudo-sentences is first embedded using BERT and then encoded using Bi-LSTM. Finally, an attention layer is used to increase the weights of the injected knowledge in the hidden representation layer of the UniLM encoder. 


\subsubsection{Pseudo-sentence embedding and encoding}
\label{subsubsec:PseudosentenceEmbeddingAndEncoding}

The basic unit of BERT embedding is a token. Thus, during the embedding phase, a pseudo-sentence set (knowledge) is first given to the BERT tokenizer. The [CLS] and [SEP] labels are added at the beginning and end of each sentence, respectively. Moreover, the tokenizer separates words into sub-words including the root of the words (e.g., ``flying'' is divided into [``fly'',``\#\#ing'']).

 There are three parts of the BERT embedding process: token embedding, position embedding, and segment embedding. Token embedding converts the tokens into vector representations in 768 dimensions ([1, n, 768], where \emph{n} is the length of a sequence). Segment embedding is used to distinguish between the keywords and target requirements in each input pair ([1,n,768]). Position embedding converts the position information in the sequence into a vector representation of 768 dimensions ([1, n, 768]). Then we sum these three embeddings and use this representation as the output of the BERT embedding.

We send the embedding to Bi-LSTM encoder \cite{DBLP:conf/icassp/GravesMH13} to obtain the pseudo-sentence encodings. The shape of the encoded hidden representation is [b,n,768], where \emph{b} is the batch size of UniLM, and \emph{n} is the sequence length.

\subsubsection{Attention mechanism}
\label{subsubsec:AttentionMechanism}

We use an attention mechanism to emphasize the importance of the injected knowledge. An attention mechanism is formally defined as follows:

\begin{small}
\begin{equation}
    \label{equ:attention}
    \begin{aligned}
        & A=softmax(QK^T/\sqrt{d}+M)\\
        & H_{ctxt}=A\cdot V \\
        & H_{knowledge}=LayerNorm(W^T\cdot H_{ctxt} + H_{UniLM}) \\
    \end{aligned}
\end{equation}
\end{small}

Here, \emph{Q} is a Query, which represents the hidden representation of the UniLM encoder layer; \emph{K} and \emph{V} are the Key and Value, respectively representing the hidden representations of the pseudo-sentences(knowledge) encoding. To obtain the attention weights of \emph{Q} and \emph{K}, we use the scaled dot product method.  $M$ is the joint mask of the pseudo-sentence and the original input of UniLM (i.e., source keywords and target specifications) 
and $d$ represents the length of the last dimension of \emph{K}. According to experience, the use of $\sqrt{d}$ reduces the sensitivity of the method to the length of \emph{K}, and improves the stability of network training. Then, we calculate the dot product between \emph{A} and \emph{V} to update \emph{V} and obtain the context representation of the attention. Finally, we sum the updated values $H_{ctxt}$ and $H_{UniLM}$, and feed them into \emph{LayerNorm} (i.e., normalization layer) to obtain the hidden representation of the knowledge.

\subsection{UniLM module}
\label{subsec:Seq2seqModule}
We modify UniLM in three ways, by adding multi-layer knowledge injection, a copy mechanism  in the training model, and a requirement syntax constrained decoding in the inference model. The knowledge injection was discussed in Sections \ref{subsec:knowledgePre} and \ref{subsec:KnowledgeInjection}. Here, we focus on the 
other two modifications.

\subsubsection{Copy mechanism}
\label{subsubsec:CopyMechanism}

Copy mechanism is very popular in NLG tasks such as translation and conversation generation \cite{DBLP:conf/acl/GuLLL16}. Its aim is to resolve the ``hard-constraint'' problem, which requires the tokens or fragments of the source sequence to occur in the target sequence. However, most copy mechanisms can only guarantee the copy of a single word, not continuous fragments \cite{kexuefm-8046}. {But in our scenario, the input keywords (phrases), even the single-term words, can be multi-term words after tokenization, making it a challenge to ensure the integrity of words in the generated statements.}

We adopt a simple but novel copy mechanism \cite{kexuefm-8046}. In our implementation, this mechanism first marks the copied fragments in the target requirements according to the source keywords. Then, we add a new copied label prediction task in the training phase of our seq2seq model to predict whether a token has been copied from the source keywords or not. In the decoder prediction stage, the original next-token prediction probability is changed to a mixed probability model of the next token prediction probability and copied token prediction probability. 





\subsubsection{Requirement syntax-constrained decoding}
\label{subsubsec:Requirementstructurizeconstraint}

Well-formed requirement specifications should follow a certain syntax, and there are several related  guidelines or models, such as EARS \cite{5328509}, IEEE 29148:2018 \cite{8559686} and M-FRDL \cite{DBLP:conf/re/GuoZL21}. They all define the fine-grained elements of single requirement specifications and each indicates one semantic role. 

The requirements generated by \emph{ReqGen} should attempt to follow at least one existing syntax so that they can accurately capture stakeholder needs \cite{8559686}. Hence, if requirement analysts follow specific requirement syntax when writing requirement specifications and they could give the semantic roles that the keywords belong to (the content in brackets of Fig. \ref{fig:in-out-example}), we would like to use this information to improve the generated sentence further. We designed an indicator called \emph{RS4RE} to evaluate the overlap of the semantic constitution of the generated requirements with those set by engineers. The \emph{RS4RE} value of each generated candidate statement is used to help select the final statement by adding it to the original probability score during beam search. 

\emph{RS4RE} is defined formally as follows. We measure the closeness of agreement of each semantic element in the generated statement and the input of an analyst. 

\begin{small}
\begin{equation}
    \label{equ:structurizer}
    \begin{aligned}
        RS4RE= \sum {\alpha_i \frac{|E_i(R_{aut}) \cap E_{i\_ref}|}{|E_{i\_ref}|}} 
        \quad where \ 1 \le i \le N
    \end{aligned}
\end{equation}
\end{small}

Let \emph{N} be the number of semantic elements in the syntax requirement analysts have selected and $E_i$ be the $i_{th}$ element. Here, $E_{i\_ref}$ is used to indicate the word set of the element $E_i$ given by the requirements analysts, and $E_i(R_{aut})$ is to indicate the set of words of the element $E_i$ in the automatically generated sentence. The agreement on element $i$ is calculated as the ratio of the scale of overlapping words to the size of the manually given set. Moreover, $\alpha_i$ is a hyper-parameter that indicates the weight of element $i$ and the sum of all $\alpha_i$ is 1.

In this work, we use M-FRDL \cite{DBLP:conf/re/GuoZL21} as the syntax because they shared the source code with us. This enables us to automatically identify the fine-grained elements in the natural language requirements, which is required by the automated calculation of \emph{RS4RE}.

\section{Experimental evaluation}
\label{sec:exp}

We evaluate \emph{ReqGen} by addressing the following research questions.


\noindent \textbf{RQ1:} How well does \emph{ReqGen} perform in comparison with existing NLG approaches on requirements specification generation based on keywords?

\noindent \textbf{RQ2}: To what extent does the multi-layer knowledge injection contribute to the requirements specification generation?

\noindent \textbf{RQ3}: To what extent does the knowledge frequency filtering contribute to the requirements specification generation?

\noindent \textbf{RQ4}: To what extent does each proposed design component contribute to the requirements specification generation?

\subsection{Experimental Design}
\label{subsec: expDesign}

\subsubsection{Data preparation}
\label{subsubsec:data}


To evaluate our \emph{ReqGen}, we use two datasets of open-source requirement specifications and two domain ontologies, i.e., Unmanned Aerial Vehicle (UAV) and Building Automation System (BAS) domains, following previous work \cite{DBLP:conf/re/ZhaoZL21}.

The UAV requirements are from the University of Notre Dame,\footnote{\url{https://dronology.info/}} including 99 requirements\cite{DBLP:conf/icse/Cleland-HuangVB18}. And the public ontology for the UAV domain\footnote{\url{http://www. dronetology.net/}} includes 400 entities. The BAS requirements are from the Standard BAS Specification (2015) \cite{BASReq} and consist of 456 requirements, involving functional, performance and security requirements. And the open-domain model of BAS\footnote{\url{https://gitlab.fi.muni.cz/ xkucer16/semanticBMS}} includes 484 entities.

The requirements are represented using natural language sentences, and we required the related keywords before the model training and testing. Thus, we performed a reverse extraction process. To be specific, we automatically extracted the noun words or phrases based on the parts-of-speech tagged by Stanford CoreNLP \cite{DBLP:conf/acl/ManningSBFBM14}. We extracted the noun and noun phrases with a series of linguistic filters for the nested noun selection, such as ${{Noun}^+{Noun}}$. For each requirement, we randomly selected \emph{n} noun phrases ($n \in [2, N]$ where \emph{N} is the total number of noun and noun phrases in one requirement) as the keywords.

To evaluate the effectiveness of the requirement syntax constrained decoding described in Section \ref{subsubsec:Requirementstructurizeconstraint}, we implemented the M-FRDL constraint \cite{DBLP:conf/re/GuoZL21} and manually assigned the semantic roles to the randomly selected keywords according to their context in requirement statements. To ensure correctness, we invited one author of the work \cite{DBLP:conf/re/GuoZL21} to check and revise them. 

Another problem with the data is that, the domain ontology may be unable to cover the all content of the requirements because these dataset were generated by different groups who have different concerns. Only if the keywords are contained in the domain ontology, will the ontology be helpful for the keyword-driven requirements generation. Thus, we performed automated domain ontology completion using an approach of \cite{https://doi.org/10.48550/arxiv.2208.06757}. {This approach first aligns the requirements and domain ontology using TransE \cite{zhong2015aligning}. Then it selects the requirement concepts, that are related to the entities in the domain ontology, and adds them as well as their corresponding relationships to the ontology.}

Given the limited data, we performed 10-fold and 5-fold cross-validation on the UAV and BAS requirements respectively. In addition, we performed a batch knowledge injection and injected the knowledge for all of the random keywords of the testing requirements, considering the potential association of the requirements. In particular, 17,681 pseudo-sentences were created for the 10 test requirements of UAV, and 139,494 pseudo-sentences for the 93 test requirements of BAS.

\subsubsection{Baselines}
\label{subsubsec: Baselines}

We selected six popular constrained text generation approaches as the baseline methods, including four pre-trained models and two other NLG models (i.e.,POINTER \cite{zhang2020pointer} and CGMH \cite{miao2019cgmh}). Similarly, we trained these models using the UAV and BAS dataset separately (10-fold and 5-fold cross-validation, respectively).
\begin{itemize}[leftmargin=1em]
    \item BERT \cite{devlin2018bert} jointly conditions on contextual information, which enables pre-training and deep bidirectional representation from unlabeled text. 
    
\item Generative Pre-trained Transformer 3(GPT3) \cite{DBLP:conf/nips/BrownMRSKDNSSAA20} uses the one-way language model training of GPT2. The model size was increased to 175 billion and 45 terabytes of data were used for training. GPT3 can perform downstream tasks without fine-tuning in a zero-shot setting. We did not perform fine-tuning neither.

\item Bidirectional and Auto-Regressive Transformers (BART) \cite{lewis2019bart} is a denoising autoencoder built with a sequence-to-sequence model suitable for various tasks. It uses a standard Transformer-based neural machine translation architecture. It is trained using text corrupted with an arbitrary noising function and by learning to reconstruct the original text.

\item UniLM \cite{DBLP:conf/nips/00040WWLWGZH19} uses three types of language modeling tasks for pre-training, which is achieved by employing a shared Transformer network and using specific self-attention masks to control the context of the prediction conditions.

\item Zhang et al. \cite{zhang2020pointer} proposed POINTER, which is based on the inserted non-autoregressive pre-training method. A beam search method was proposed to achieve log-level non-autoregressive generation.

\item The Constrained Generation by Metropolis-Hastings sampling (CGMH) method \cite{miao2019cgmh} can cope with both hard and soft constraints during sentence generation. Different from the traditional latent space usage, it directly samples from the sentence space using the Metropolis-Hastings sampling.
\end{itemize}

\subsubsection{Metrics}
\label{subsubsec:metrics}

We selected \emph{BLEU} \cite{papineni2002bleu} as the first metric, which is commonly used in machine translation, NLG and source code generation. It measures the degree of overlap between the generated and reference sentences using n-grams. The higher the degree of overlap, the higher the quality of the generated text. Here, \emph{BLEU1} measures word-level accuracy and \emph{BLEU2} measures sentence fluency to a specific degree. 
In addition, we employed the \emph{ROUGE} metric \cite{lin2004rouge}. 
\emph{ROUGE-N} calculates the total sum of the number of n-grams occurring in both the generated and target sentences, and \emph{ROUGE-L} calculates the longest common subsequence. We calculated recall, precision and F-measure for each kind of ROUGE rather than the simple n-gram recall\cite{lin2004rouge}. 



Besides, we used the \emph{RS4RE} metric to evaluate the agreement of syntax compliance between the generated and target sentences, as described in Section \ref{subsubsec:Requirementstructurizeconstraint}. Here, $E_{i\_ref}$ refers to the set of words in the $i_{th}$ semantic element of the target requirement. We also recorded the time used by each model (except GPT-3) for training and testing, considering its practical value.

\subsection{Experimental results and analysis}
\label{subsec: results}

\subsubsection{\textbf{RQ1 effectiveness of our approach}}

We illustrate the experimental results for UAV and BAS cases in Table \ref{tab:baseline}, and Table \ref{tab:baselineexample} presents an example of the requirement generated by each baseline and \emph{ReqGen}, from which we can draw three conclusions:

(1) \textbf{Our method achieves the best or second-best results for all metrics.} 
In the UAV case, our method yields the best results for five metrics and the second-best results for the remaining eight metrics. BART achieves good performance with six best and five second-best results, which demonstrates the benefits of its hard constraint design. However, it can only embed a single input word and loses the complete semantic meaning of phrases (i.e., it obtains better BLEU1 and ROUGE-1 results, but weaker BLEU2 and ROUGE-2 results), which can also be seen in the example in Table \ref{tab:baselineexample}. Moreover, \emph{ReqGen} obtains a better RS4RE result than BART, indicating that it has a better semantic-oriented text generation ability.

On the BAS domain, our \emph{ReqGen} achieves seven best and four second-best results, outperforming all baselines. The basic UniLM performs moderately well because it considers both the input and its context during the next token prediction. This is the reason that we select it as the backbone of \emph{ReqGen}. 


(2) \textbf{The four pre-trained models perform better than the other two baselines.} Among the four pre-trained models in the UAV domain, BART obtains the best ROUGE-1, ROUGE-2, and ROUGE-L results. It even outperforms \emph{ReqGen} in ROUGE-1, which is possible because BART forces all keywords to be included in the output during decoding. However, because it only sees the input, it performs worse in ROUGE-2. UniLM obtains good BLEU1 and BLEU2 results and the best ROUGE-2 results in the BAS case, indicating its better ability to generate relatively fluent statements. 

CGMH performs worst according to the two BLEU metrics and the F-measure of the three ROUGE metrics, as shown in Tables \ref{tab:baseline} and \ref{tab:baselineexample}. POINTER also has a weak performance, and a possible reason for this performance could be that it places the entire generation burden on the decoder, which means that the sentences it generates are very long, but the correlation with the target is weak (see the example in Table \ref{tab:baselineexample}).

(3) \textbf{Our method is slightly slower than UniLM and BART.} As shown in the last column of 
Table \ref{tab:baseline}, we recorded the average time consumed by each method (in hours) to perform the 10-fold cross-validation on the UAV and 5-fold cross-validation on the BAS data. We did not record the time used by GPT3 because it does not need extra fine-tuning \cite{DBLP:conf/nips/BrownMRSKDNSSAA20}. We observe that \emph{ReqGen} needs slightly more time than UniLM and BART, but less time than the other approaches, even although it uses extra knowledge injection, a copy mechanism, and requirement syntax constrained decoding. 

\begin{table*}[!htbp]
\centering
\caption{Results of the baselines and our \emph{ReqGen}}
\label{tab:baseline}
\begin{tabular}{|c|c|c|c|ccc|ccc|ccc|c|c|}
\hline
\multirow{2}{*}{}    & \multirow{2}{*}{Method} & \multirow{2}{*}{BLEU1} & \multirow{2}{*}{BLEU2} & \multicolumn{3}{c|}{ROUGE-1}                                        & \multicolumn{3}{c|}{ROUGE-2}                                        & \multicolumn{3}{c|}{ROUGE-L} & \multirow{2}{*}{RS4RE} & \multirow{2}{*}{\begin{tabular}[c]{@{}c@{}}Time\\ (Hours)\end{tabular}}  \\ \cline{5-13} 
                     &                         &                        &                       & \multicolumn{1}{c}{R.} & \multicolumn{1}{c}{P.} & F.     & \multicolumn{1}{c}{R.} & \multicolumn{1}{c}{P.} & F.     & \multicolumn{1}{c}{R.} & \multicolumn{1}{c}{P.} & F.     &     &                   \\ \hline
\multirow{7}{*}{UAV} & Bert\_base              & 37.92                  & 23.53                 & \multicolumn{1}{c}{54.33}  & \multicolumn{1}{c}{42.04}     & 47.40 & \multicolumn{1}{c}{31.32}  & \multicolumn{1}{c}{24.88}     & 27.73 & \multicolumn{1}{c}{51.29}  & \multicolumn{1}{c}{39.90}     & 44.88 &8.00 & 1.5\\ \cline{2-15} 
                     & GPT3                     & 34.20                  & 18.28                 & \multicolumn{1}{c}{34.51}  & \multicolumn{1}{c}{\textbf{59.42}}     & 43.66 & \multicolumn{1}{c}{16.49}  & \multicolumn{1}{c}{14.82}     & 15.61 & \multicolumn{1}{c}{31.02}  & \multicolumn{1}{c}{56.22}     & 39.98 & 1.31 & -\\ \cline{2-15} 
                     & UniLM                   & 34.27                  & 17.83                 & \multicolumn{1}{c}{55.15}  & \multicolumn{1}{c}{41.53}     & 47.38 & \multicolumn{1}{c}{18.21}  & \multicolumn{1}{c}{14.44}     & 16.11 & \multicolumn{1}{c}{49.73}  & \multicolumn{1}{c}{37.71}     & 42.89 & 2.54 & 1\\ \cline{2-15} 
                     & BART                    & \textbf{42.66}                  & 23.99                 & \multicolumn{1}{c}{\textbf{71.75}}  & \multicolumn{1}{c}{52.12}     & \textbf{60.38} & \multicolumn{1}{c}{37.86}  & \multicolumn{1}{c}{26.90}     & 31.45 & \multicolumn{1}{c}{\textbf{66.07}}  & \multicolumn{1}{c}{\textbf{47.79}}     & \textbf{55.47} & 5.76 & 1\\ \cline{2-15} 
                     & CGMH                    & 12.07                  & 1.84                  & \multicolumn{1}{c}{35.06}  & \multicolumn{1}{c}{17.86}     & 23.66 & \multicolumn{1}{c}{5.48}   & \multicolumn{1}{c}{2.47}      & 3.41  & \multicolumn{1}{c}{32.42}  & \multicolumn{1}{c}{16.52}     & 21.89 & 0.00 & 22\\ \cline{2-15} 
                     & POINTER                 & 17.26                  & 2.46                  & \multicolumn{1}{c}{24.15}  & \multicolumn{1}{c}{38.66}     & 29.73 & \multicolumn{1}{c}{2.83}   & \multicolumn{1}{c}{5.20}      & 3.67  & \multicolumn{1}{c}{19.92}  & \multicolumn{1}{c}{32.07}     & 24.58 & 4.60 & 1\\ \cline{2-15} 
                     & ReqGen                     & 42.15                  & \textbf{25.04}                 & \multicolumn{1}{c}{69.93}  & \multicolumn{1}{c}{49.91}     & 58.25 & \multicolumn{1}{c}{\textbf{39.03}}  & \multicolumn{1}{c}{\textbf{28.21}}     & \textbf{32.75} & \multicolumn{1}{c}{65.12}  & \multicolumn{1}{c}{46.47}     & 54.23 & \textbf{8.89} & 1.2 \\ \hline
\multirow{7}{*}{BAS} & Bert\_base              & 29.08                  & 9.28                  & \multicolumn{1}{c}{46.69}  & \multicolumn{1}{c}{35.90}     & 40.59 & \multicolumn{1}{c}{12.84}  & \multicolumn{1}{c}{10.57}     & 11.60 & \multicolumn{1}{c}{40.79}  & \multicolumn{1}{c}{31.64}     & 35.64 & 12.48 & 1 \\ \cline{2-15} 
                     & GPT3                     & 28.34                  & 7.87                  & \multicolumn{1}{c}{37.17}  & \multicolumn{1}{c}{41.07}     & 33.56 & \multicolumn{1}{c}{10.02}  & \multicolumn{1}{c}{9.64}      & 9.30 & \multicolumn{1}{c}{31.98}  & \multicolumn{1}{c}{35.74}     & 30.53 & 7.08 & -\\ \cline{2-15} 
                     & UniLM                   & 30.08                  & 13.31                 & \multicolumn{1}{c}{42.44}  & \multicolumn{1}{c}{33.58}     & 37.49 & \multicolumn{1}{c}{\textbf{24.42}}  & \multicolumn{1}{c}{\textbf{18.86}}     & \textbf{21.28} & \multicolumn{1}{c}{39.58}  & \multicolumn{1}{c}{31.47}     & 35.06 & 13.56 & 0.8\\ \cline{2-15} 
                     & BART                    & 27.05                  & 10.60                 & \multicolumn{1}{c}{\textbf{63.67}}  & \multicolumn{1}{c}{37.37}     & 47.10 & \multicolumn{1}{c}{20.59}  & \multicolumn{1}{c}{12.43}     & 15.50 & \multicolumn{1}{c}{\textbf{58.78}}  & \multicolumn{1}{c}{34.14}     & 43.19 & 14.58 & 0.7\\ \cline{2-15} 
                     & CGMH                    & 10.58                       & 1.85                     & \multicolumn{1}{c}{57.69}       & \multicolumn{1}{c}{20.94}          & 30.73      & \multicolumn{1}{c}{4.23}       & \multicolumn{1}{c}{1.64}          & 2.36      &  \multicolumn{1}{c}{44.70}       & \multicolumn{1}{c}{16.41}          & 24.01 & 3.26 & 72     \\ \cline{2-15} 
                     & POINTER                 & 21.34                  & 2.34                  &  \multicolumn{1}{c}{25.34}  & \multicolumn{1}{c}{43.82}     & 32.11 & \multicolumn{1}{c}{1.93}   & \multicolumn{1}{c}{3.69}      & 2.53  & \multicolumn{1}{c}{21.01}  & \multicolumn{1}{c}{36.53}     & 26.68 & 5.72 & 1 \\ \cline{2-15} 
                     & ReqGen                     & \textbf{38.07}                  & \textbf{15.62}                 & \multicolumn{1}{c}{58.96}  & \multicolumn{1}{c}{\textbf{44.63}}     & \textbf{50.80} & \multicolumn{1}{c}{20.35}  & \multicolumn{1}{c}{16.45}     & 18.19 & \multicolumn{1}{c}{52.82}  & \multicolumn{1}{c}{\textbf{40.17}}     & \textbf{45.63} & \textbf{17.42} & 0.9\\ \hline
\end{tabular}
\end{table*}

\begin{table}[]
\centering
	\caption{Examples showing the requirement generated by the baselines and our method from three keywords.}
	\label{tab:baselineexample}
	\footnotesize
	\begin{tabular}{p{1cm}|p{7cm}}
		\hline
		keywords     & landing, internal simulator, ground  \\ \hline
		Target     & When given a \textbf{landing} command the \textbf{internal} \textbf{simulator} shall move  the UAV from to the \textbf{ground} altitude corresponding to its current  longitude and latitude. \\ \hline
	
		Bert\_base & when a flight \textbf{simulator} is activated the  flight \textbf{simulator} shall compute the  location of the uav.                                                         \\ \hline
		GPT3       & When a UAV lands the \textbf{Internal Simulator} shall record the time of \textbf{landing}.    \\ \hline
		
		UniLM     & When a UAV is loaded the \textbf{internal} \textbf{simulator} shall display the location of the UAV.  \\ \hline
		
		BART & \textbf{landing} \textbf{internal} flight \textbf{simulator} shall compute the \textbf{ground} position of a UAV. \\ \hline
		
		CGMH       & \textbf{simulator} owners for takeoff at a corresponding latitude. \\ \hline
		
		POINTER    & in which a single point \textbf{landing} system, when a single point that is assigned a separate gps system or for an active navigation system has to orient at its current position and to display the current \textbf{ground} coordinates. \\ \hline
		
		ReqGen        & When a \textbf{landing} UAV is assigned to a UAV, the  \textbf{internal} \textbf{simulator} shall compute the  \textbf{ground}  longitude latitude and latitude of the UAV.    \\ \hline
		
	\end{tabular}
\end{table}

\subsubsection{\textbf{RQ2 effectiveness of multi-layer knowledge injection}}
\label{sec:exptimesofmultiinjection}

We also evaluated the effectiveness of the multi-layer knowledge injection in the two domains. To implement and evaluate \emph{ReqGen}, given that we collected 5-hop keyword-related information and there are 12 layers in UniLM, we had to determine 1) \emph{which levels should have knowledge injected}, and 2) \emph{what kind of knowledge should be injected into the different layers.} Because of the uncertainty in the injected layers and the injected knowledge for one specific layer (i.e., which hop in the five hops), there are many combinations that could have been used in our experiments. Thus, we pruned the candidate combinations for the experiments.

As for the injected layers, Li et al. \cite{DBLP:journals/corr/abs-2012-10813} injected knowledge into the $1^{st}$ layer. Jawahar et al. \cite{DBLP:conf/acl/JawaharSS19} experimented with different layers of BERT on 10 sentence-level detection tasks. They observed that the $1^{st}$ and $2^{nd}$ layers learn surface information such as word detection in sentences. Layers 4 to 7 learn syntactic information such as word order sensitivity. Layers 8 to 12 learn semantic-level information such as subject-verb agreement. Inspired by these two studies, we focused on three combinations of shallow layers, shallow + syntactic layers, shallow + syntactic + semantic layers, with layer combinations of (1,2),(1,2,4) and (1,2,4,8), respectively. Moreover, considering the five hops in our collected information, we made an extra evaluation on the combination (1,2,4,8,11) and each layer was assigned 1-hop information.

For the injected knowledge, we followed the cognitive process of human learning, in which human usually start from general and overview knowledge and then pay close attention to critical parts. Similarly, we assigned more keyword-knowledge in the lower layers. For example, for the layer combination (1,2), we assigned all 5-hop knowledge to the first layer and the most related 1-hop knowledge to the second layer. The injected knowledge details of different layer combinations can be found in the \emph{layer(hop)} of Table \ref{tab:timesofmultiinjection}, which presents the results of this experiment.

For the UAV case, we observe that the best BLEU1 and BLEU2 scores are achieved by the (1,2,4) combination. For ROUGE-1, the best recall is achieved by injecting knowledge into layers 1, 2, 4 and 8, the best precision is achieved by the (1,2) combination, and the best of F-measure is achieved by the (1,2,4) combination. Similar phenomenon are observed for the ROUGE-2 and ROUGE-L metrics. In the BAS case, the best BLEU1 and BLEU2 results are also achieved by the (1,2,4) combination. For ROUGE-1, the best F-value is achieved by the (1,2) combination; however, for ROUGE-2 and ROUGE-L, the best performance is achieved by the (1,2,4) combination. 

In summary, the (1,2,4) combination with the 5-, 2-, and 1-hop knowledge injection achieves seven best results in the UAV case, and nine best results in the BAS case, out of all 11 metrics. Hence, we suggest three knowledge injections by injecting all 5-hop knowledge into the first layer, 2-hop knowledge into the second layer, and the most critical 1-hop knowledge into the fourth layer.


\begin{table*}[!htbp]
\centering
\caption{The effectiveness of different injection times}
\label{tab:timesofmultiinjection}
\begin{tabular}{|c|c|c|c|ccc|ccc|ccc|}
\hline
\multirow{2}{*}{}    & \multirow{2}{*}{Layer(hop)} & \multirow{2}{*}{BLEU1} & \multirow{2}{*}{BLEU2} & \multicolumn{3}{c|}{ROUGE-1}                                        & \multicolumn{3}{c|}{ROUGE-2}                                        & \multicolumn{3}{c|}{ROUGE-L}                                        \\ \cline{5-13} 
                     &                        &                        &                       & \multicolumn{1}{c}{R.} & \multicolumn{1}{c}{P.} & F.     & \multicolumn{1}{c}{R.} & \multicolumn{1}{c}{P.} & F.     & \multicolumn{1}{c}{R.} & \multicolumn{1}{c}{P.} & F.     \\ \hline
\multirow{4}{*}{UAV} & 1(5),2(1)                    & 43.23                  & 24.63                 & \multicolumn{1}{c}{63.21}  & \multicolumn{1}{c}{50.51}     & 56.15
 & \multicolumn{1}{c}{34.90}  & \multicolumn{1}{c}{\textbf{27.61}}     & 29.86 & \multicolumn{1}{c}{58.83}  & \multicolumn{1}{c}{\textbf{47.04}}     & 51.12 \\ \cline{2-13} 
                     & 1(5),2(2),4(1)                  & \textbf{43.87}                  & \textbf{25.05}                 & \multicolumn{1}{c}{65.69}  & \multicolumn{1}{c}{50.34}     & \textbf{57.00} & \multicolumn{1}{c}{\textbf{35.59}}  & \multicolumn{1}{c}{27.18}     & \textbf{30.82} & \multicolumn{1}{c}{61.13}  & \multicolumn{1}{c}{46.86}     & \textbf{53.05} \\ \cline{2-13} 
                     & 1(5),2(3),4(2),8(1)                & 42.57                  & 23.63                 & \multicolumn{1}{c}{\textbf{66.22}}  & \multicolumn{1}{c}{49.96}     & 56.95 & \multicolumn{1}{c}{35.14}  & \multicolumn{1}{c}{26.10}     & 29.95 & \multicolumn{1}{c}{\textbf{61.39}}  & \multicolumn{1}{c}{46.57}     & 52.95 \\ \cline{2-13} 
                     & 1(5),2(4),4(3),8(2),11(1)             & 40.2                   & 21.64                 & \multicolumn{1}{c}{63.04}  & \multicolumn{1}{c}{47.37}     & 54.09 & \multicolumn{1}{c}{32.40}  & \multicolumn{1}{c}{24.17}     & 26.89 & \multicolumn{1}{c}{58.65}  & \multicolumn{1}{c}{44.17}     & 49.42 \\ \hline
                    
\multirow{4}{*}{BAS} & 1(5),2(1)                    & 34.51                  & 13.25                 & \multicolumn{1}{c}{\textbf{55.54}}  & \multicolumn{1}{c}{41.74}     & \textbf{47.66} & \multicolumn{1}{c}{17.86}  & \multicolumn{1}{c}{14.13}     & 15.78 & \multicolumn{1}{c}{49.87}  & \multicolumn{1}{c}{37.66}     & 42.91 \\ \cline{2-13} 
                     & 1(5),2(2),4(1)                  & \textbf{34.55}                  & \textbf{13.73}                 & \multicolumn{1}{c}{55.02}  & \multicolumn{1}{c}{\textbf{41.76}}     & 47.48 & \multicolumn{1}{c}{\textbf{18.30}}  & \multicolumn{1}{c}{\textbf{14.68}}     & \textbf{16.29} & \multicolumn{1}{c}{\textbf{49.83}}  & \multicolumn{1}{c}{\textbf{38.03}}     & \textbf{43.14} \\ \cline{2-13} 
                     & 1(5),2(3),4(2),8(1)                & 33.26                  & 12.92                & \multicolumn{1}{c}{54.51}  & \multicolumn{1}{c}{40.64}     & 46.56 & \multicolumn{1}{c}{18.13}  & \multicolumn{1}{c}{14.19}     & 15.92 & \multicolumn{1}{c}{49.21}  & \multicolumn{1}{c}{36.92}     & 42.19 \\ \cline{2-13} 
                     & 1(5),2(4),4(3),8(2),11(1)             & 32.33                  & 12.31                 & \multicolumn{1}{c}{54.27}  & \multicolumn{1}{c}{40.16}     & 46.16 & \multicolumn{1}{c}{17.38}  & \multicolumn{1}{c}{13.77}     & 15.37 & \multicolumn{1}{c}{49.02}  & \multicolumn{1}{c}{36.55}     & 41.88 \\ \hline 
\end{tabular}
\end{table*}

\subsubsection{\textbf{RQ3 effectiveness of frequency filtering in the knowledge search}}
\label{subsubsec:frequencyFiltering}

We further evaluated the impact of different frequency sets in the knowledge search on the final requirement generation (as described in Section \ref{subsubsec: multi-hopSearching}). We regarded more frequent knowledge as more important, and our aim was to inject only important information into \emph{ReqGen} to reduce noise pollution in the data.

To evaluate the impact of a frequency filter (a single dependent variable), we set the injected layer as a constant. We experimented with frequency thresholds of 0, 10, and 50 on all 5-hop knowledge injected into the $1^{st}$ layer. In other words, we injected all traversed entities, entities occurring more than 10 times, or those occurring more than 50 times in the ontology graph traversal. The results are shown in Table \ref{tab:frequencyfiltering} and the best result of each metric is highlighted with bold font.

We observe different results for the two cases. For UAV, no frequency filtering is best, followed by 10- and 50- frequency filters. Whereas, for BAS, the best performance is achieved with the 10-frequency filter, and 50-frequency filter is the second best. This result leads us to the following two observations. (1) Intuitively, the frequency set is strongly correlated with the scale of injected knowledge. In our case, the 5-hop knowledge of the UAV domain includes 17,681 pseudo-sentences and BAS includes 139,494 sentences (approximately 7.9 times). (2) When there is a massive amount of information, less and more refined knowledge is more valuable. For example, 50-frequency filter is better than no filter in the BAS domain. However, when the information is small, all related knowledge is valuable (e.g., in the UAV case, no filter is better than the 10-frequency filter, which is better than the 50-frequency filter). However, because of the limited available cases, we cannot give a criterion for frequency filter selection temporarily in this initial study.



\begin{table*}[!h]
\centering
\caption{The effectiveness of knowledge frequency filtering}
\label{tab:frequencyfiltering}
\begin{tabular}{|c|c|c|c|ccc|ccc|ccc|}
\hline
\multirow{2}{*}{}    & \multirow{2}{*}{\begin{tabular}[c]{@{}c@{}}Frequency  filtering\end{tabular}} & \multirow{2}{*}{BLEU1} & \multirow{2}{*}{BLEU2} & \multicolumn{3}{c|}{ROUGE-1}                                                             & \multicolumn{3}{c|}{ROUGE-2}                                                             & \multicolumn{3}{c|}{ROUGE-L}                                                             \\ \cline{5-13} 
                     &                                                                                &                        &                       & \multicolumn{1}{c}{R.} & \multicolumn{1}{c}{P.} & F.                          & \multicolumn{1}{c}{R.} & \multicolumn{1}{c}{P.} & F.                          & \multicolumn{1}{c}{R.} & \multicolumn{1}{c}{P.} & F.                          \\ \hline
\multirow{3}{*}{UAV} & no                                                                             & 42.78                  & \textbf{24.14}                 & \multicolumn{1}{c}{\textbf{66.54}}  & \multicolumn{1}{c}{49.85}     & \textbf{57.00}                      & \multicolumn{1}{c}{\textbf{35.51}}  & \multicolumn{1}{c}{\textbf{26.59}}     & \textbf{30.41}                      & \multicolumn{1}{c}{\textbf{61.76}}  & \multicolumn{1}{c}{46.45}     & \textbf{53.02}                      \\ \cline{2-13} 
                     & 10                                                                             & \textbf{43.50}                  & 23.97                 & \multicolumn{1}{c}{65.80}  & \multicolumn{1}{c}{\textbf{49.92}}     & 56.77                      & \multicolumn{1}{c}{34.99}  & \multicolumn{1}{c}{26.22}     & 29.24                      & \multicolumn{1}{c}{61.75}  & \multicolumn{1}{c}{\textbf{47.04}}     & 52.46                      \\ \cline{2-13} 
                     & 50                                                                             & 40.47                  & 21.39                 & \multicolumn{1}{c}{61.87}  & \multicolumn{1}{c}{46.58}     & 53.15                      & \multicolumn{1}{c}{32.05}  & \multicolumn{1}{c}{23.63}     & 27.21                      & \multicolumn{1}{c}{57.90}  & \multicolumn{1}{c}{43.70}     & 49.80                      \\ \hline
\multirow{3}{*}{BAS} & no                                                                             & 34.15                  & 13.37                 & \multicolumn{1}{c}{54.75}  & \multicolumn{1}{c}{41.30}     & \multicolumn{1}{l|}{47.08} & \multicolumn{1}{c}{18.01}  & \multicolumn{1}{c}{14.49}     & \multicolumn{1}{l|}{16.06} & \multicolumn{1}{c}{49.41}  & \multicolumn{1}{c}{37.52}     & \multicolumn{1}{l|}{42.65} \\ \cline{2-13} 
                     & 10                                                                             & \textbf{34.43}                  & \textbf{13.65}                 & \multicolumn{1}{c}{\textbf{55.93}}  & \multicolumn{1}{c}{\textbf{41.76}}     & \multicolumn{1}{l|}{\textbf{47.82}} & \multicolumn{1}{c}{\textbf{18.84}}  & \multicolumn{1}{c}{\textbf{14.80}}     & \multicolumn{1}{l|}{\textbf{16.58}} & \multicolumn{1}{c}{\textbf{50.29}}  & \multicolumn{1}{c}{37.76}     & \multicolumn{1}{l|}{43.13} \\ \cline{2-13} 
                     & 50                                                                             & 34.33                  & 13.45                 & \multicolumn{1}{c}{55.68}  & \multicolumn{1}{c}{41.69}     & \multicolumn{1}{l|}{47.68} & \multicolumn{1}{c}{18.65}  & \multicolumn{1}{c}{14.66}     & \multicolumn{1}{l|}{16.41} & \multicolumn{1}{c}{50.28}  & \multicolumn{1}{c}{\textbf{37.83}}     & \multicolumn{1}{l|}{\textbf{43.18}} \\ \hline
\end{tabular}
\end{table*}

\subsubsection{\textbf{RQ4 ablation experiment}}

The ablation experiment was designed to verify the effectiveness of each critical proposed component in our \emph{ReqGen}: multi-layer injection (including frequency filtering), copy mechanism and syntax-constrained decoding. Table \ref{tab:ablationexperiment} shows the results of \emph{ReqGen} and its four variants. For the convenience of comparison, we indicate the values, that are better than those in the previous row with an up-arrow.

In the UAV case, compared with the one layer knowledge injection, the three injections into layers of 1, 2 and 4 yield an approximately 1\% increase in BLEU1 and BLEU2. For the ROUGE results, the improvement is primarily in the three precision scores, which are increased by 0.5\%. However, the recall in ROUGE was slightly decreased, possibly due to some related knowledge loss in layers 2 and 4. However, the F-measure values remained unchanged or turn better with respect to the values obtained for the single-layer injection. In the BAS case, all metrics, including the BLEU and three ROUGE metrics, were improved by three-layer knowledge injection. This observation shows that the multi-layer injection design mainly improves the BLEU metric and the precision of ROUGE, indicating that this mechanism enhances the ratio of valid words in the final statements. In other words, although the single-layer injection can help the model to obtain the knowledge, \emph{the multi-layer injection is further helpful for knowledge absorption}.


\begin{table*}[!h]
\centering
\caption{The ablation experiment}
\label{tab:ablationexperiment}
\begin{tabular}{|c|c|c|c|ccc|ccc|ccc|}
\hline
\multirow{2}{*}{}    & \multirow{2}{*}{\begin{tabular}[c]{@{}c@{}}Settings\end{tabular}}                 & \multirow{2}{*}{BLEU1} & \multirow{2}{*}{BLEU2} & \multicolumn{3}{c|}{ROUGE-1}                                        & \multicolumn{3}{c|}{ROUGE-2}                                        & \multicolumn{3}{c|}{ROUGE-L}                                        \\ \cline{5-13} 
                     &                                                                                                &                        &                       & \multicolumn{1}{c}{R.} & \multicolumn{1}{c}{P.} & F.     & \multicolumn{1}{c}{R.} & \multicolumn{1}{c}{P.} & F.     & \multicolumn{1}{c}{R.} & \multicolumn{1}{c}{P.} & F.     \\ \hline
\multirow{5}{*}{UAV} & Layer 1                                                                                        & 42.78                  & 24.14                 & \multicolumn{1}{c}{66.55}  & \multicolumn{1}{c}{49.86}     & 57.00 & \multicolumn{1}{c}{35.51}  & \multicolumn{1}{c}{26.59}     & 29.62 & \multicolumn{1}{c}{61.76}  & \multicolumn{1}{c}{46.45}     & 53.02 \\ \cline{2-13} 

                     & Layer 1,2,4                                                                                    & 43.87 $\uparrow$                 & 25.05$\uparrow$                 & \multicolumn{1}{c}{65.69}  & \multicolumn{1}{c}{50.34$\uparrow$}     & 57.00 & \multicolumn{1}{c}{35.59$\uparrow$}  & \multicolumn{1}{c}{27.18$\uparrow$}     & 30.82$\uparrow$ & \multicolumn{1}{c}{61.13}  & \multicolumn{1}{c}{46.86$\uparrow$}     & 53.05$\uparrow$ \\ \cline{2-13} 
                     
                     & \begin{tabular}[c]{@{}c@{}}Layer 1,2,4+ 10 Fre.\end{tabular}                                  & \textbf{44.80}$\uparrow$                  & 25.12$\uparrow$                 & \multicolumn{1}{c}{65.35}  & \multicolumn{1}{c}{\textbf{52.18}$\uparrow$}     & 58.03$\uparrow$ & \multicolumn{1}{c}{35.48}  & \multicolumn{1}{c}{28.11$\uparrow$}     & 31.37$\uparrow$ & \multicolumn{1}{c}{60.95}  & \multicolumn{1}{c}{\textbf{48.93}$\uparrow$}     & \textbf{54.28}$\uparrow$ \\ \cline{2-13} 
                     
                     & \begin{tabular}[c]{@{}c@{}}Layer 1,2,4+\\ 10 Fre.+Copy\end{tabular}                             & 42.15                  & 25.04                 & \multicolumn{1}{c}{\textbf{69.93}$\uparrow$}  & \multicolumn{1}{c}{49.91}     & \textbf{58.25}$\uparrow$ & \multicolumn{1}{c}{\textbf{39.03}$\uparrow$}  & \multicolumn{1}{c}{\textbf{28.21}$\uparrow$}     & \textbf{32.75}$\uparrow$ & \multicolumn{1}{c}{\textbf{65.12}$\uparrow$}  & \multicolumn{1}{c}{46.47}     & 54.23 \\ \cline{2-13} 
                     
                     & \begin{tabular}[c]{@{}c@{}}Layer 1,2,4+ 10 Fre.\\+Copy+ Syntax \\cons. (\emph{ReqGen})\end{tabular} & 42.15                  & 25.04                 & \multicolumn{1}{c}{\textbf{69.93}}  & \multicolumn{1}{c}{49.91}     & \textbf{58.25} & \multicolumn{1}{c}{\textbf{39.03}}  & \multicolumn{1}{c}{\textbf{28.21}}     & \textbf{32.75} & \multicolumn{1}{c}{\textbf{65.12}}  & \multicolumn{1}{c}{46.47}     & 54.23 \\ \hline
\multirow{5}{*}{BAS} & Layer 1                                                                                        & 34.15                  & 13.37                 & \multicolumn{1}{c}{54.75}  & \multicolumn{1}{c}{41.30}     & 47.08 & \multicolumn{1}{c}{18.01}  & \multicolumn{1}{c}{14.49}     & 16.06 & \multicolumn{1}{c}{49.41}  & \multicolumn{1}{c}{37.52}     & 42.65 \\ \cline{2-13} 

                     & Layer 1,2,4                                                                                    & 34.55$\uparrow$                  & 13.73 $\uparrow$                & \multicolumn{1}{c}{55.02$\uparrow$}  & \multicolumn{1}{c}{41.76$\uparrow$}     & 47.48$\uparrow$ & \multicolumn{1}{c}{18.30$\uparrow$}  & \multicolumn{1}{c}{14.68$\uparrow$}     & 16.29$\uparrow$ & \multicolumn{1}{c}{49.83$\uparrow$}  & \multicolumn{1}{c}{38.03$\uparrow$}     & 43.14$\uparrow$ \\ \cline{2-13} 
                     
                     & \begin{tabular}[c]{@{}c@{}}Layer 1,2,4+ 10 Fre.\end{tabular}                                  & 37.99$\uparrow$                  & 15.57$\uparrow$                 & \multicolumn{1}{c|}{56.30$\uparrow$}  & \multicolumn{1}{c}{\textbf{45.05}$\uparrow$}     & 50.05$\uparrow$ & \multicolumn{1}{c}{19.58$\uparrow$}  & \multicolumn{1}{c}{\textbf{16.57}$\uparrow$}     & 17.95$\uparrow$ & \multicolumn{1}{c}{50.60$\uparrow$}  & \multicolumn{1}{c}{\textbf{40.68}$\uparrow$}     & 45.10$\uparrow$ \\ \cline{2-13} 
                     
                     & \begin{tabular}[c]{@{}c@{}}Layer 1,2,4+\\ 10 Fre.+Copy\end{tabular}                             & \textbf{38.41}$\uparrow$                  & 15.06                 & \multicolumn{1}{c}{\textbf{59.47}$\uparrow$}  & \multicolumn{1}{c}{44.93}     & \textbf{51.18$\uparrow$} & \multicolumn{1}{c}{19.46}  & \multicolumn{1}{c}{15.75}     & 17.41 & \multicolumn{1}{c}{\textbf{53.30}$\uparrow$ }  & \multicolumn{1}{c}{40.48}     & \textbf{46.01$\uparrow$ } \\ \cline{2-13} 
                     
                     & \begin{tabular}[c]{@{}c@{}}Layer 1,2,4+ 10 Fre.\\+Copy+ Syntax \\cons. (\emph{ReqGen})\end{tabular} & 38.07                  & \textbf{15.62} $\uparrow$  & \multicolumn{1}{c}{58.96}  & \multicolumn{1}{c}{44.63}     & 50.80 & \multicolumn{1}{c}{\textbf{20.35}$\uparrow$ }  & \multicolumn{1}{c}{16.45$\uparrow$ }     & \textbf{18.19$\uparrow$ } & \multicolumn{1}{c}{52.82}  & \multicolumn{1}{c}{40.17}     & 45.63 \\ \hline
\end{tabular}
\end{table*}

The comparisons between the second and third rows in both the UAV and BAS domains show the impact of 10-frequency filtering on layers of 1, 2, and 4. We can see that the two BLEU values in both cases are better with this frequency setting. For UAV, the precision and F-measure of the three ROUGE metrics have been increased further. However, the recall values are weaker. This is expected because the filtering operation on a relatively small set of knowledge would probably lessen the number of useful clues, whereas, for BAS, due to its larger scale, all metrics have been improved even after filtering. We note that  \emph{the 10-frequency filtering indeed effectively reduces the noise}.


\begin{figure}[!htbp]
\centering
 \includegraphics[trim={0 12cm 0 0}, clip, width=0.49\textwidth]{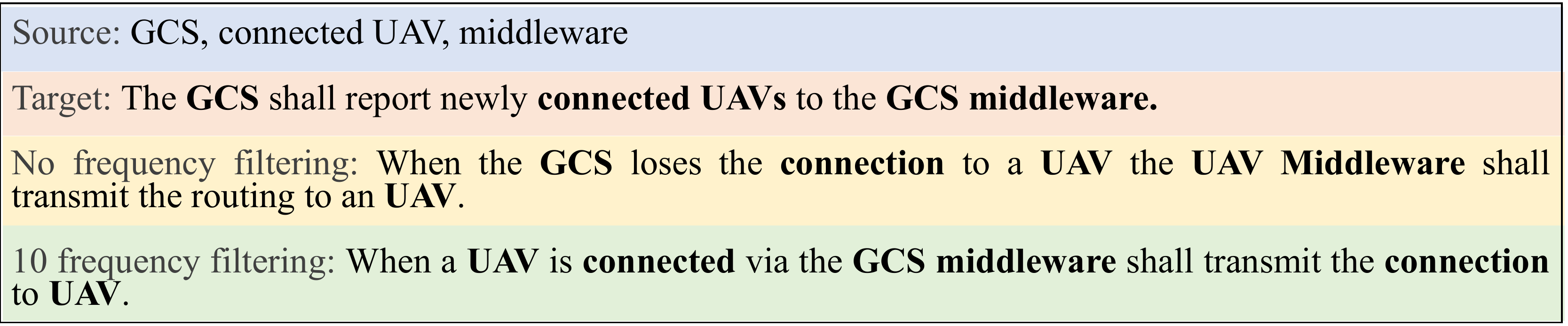} 
\caption{An example with or without frequency filtering}\label{fig:exp4frefiltering}
\end{figure}


Fig \ref{fig:exp4frefiltering} shows one example of the effects of frequency filtering. In this example, when no-frequency filtering is used, the terms ``loses'' and ``routing'' in the domain ontology are injected into \emph{ReqGen} and reflected in the generated requirement, showing that noise words are reflected in the lexical constitutions of the generated requirements, which then determine the semantic meaning. By contrast, the example with 10-frequency filtering shows that our frequency filtering method helps reducing the noise in the injected knowledge.
 
The comparisons between the fourth and fifth rows of Table \ref{tab:ablationexperiment} for both domains show the influence of the copy mechanism. From the results for the UAV domain, we observe that six results, mainly ROUGE scores, have been improved and achieve the best values with this design, whereas the two BLEU scores have fallen approximately 2\%. Similarly, the results for five metrics have been increased and the remaining six are decreased for the BAS domain. This shows that the copy mechanism is helpful for the \emph{recall of the overlapping n-grams} in the generated sentences. However, because it is designed to guarantee that the given keywords must appear in the generated sentence, this actually increases the length of the generated sentences. In other words, it is highly likely to dilute the ratio of the overlapping n-grams in the whole statements meanwhile, which decreases the BLEU scores.

\begin{figure}[!htbp]
\centering
 \includegraphics[trim={0 13.5cm 0.5cm 0}, clip, width=0.49\textwidth]{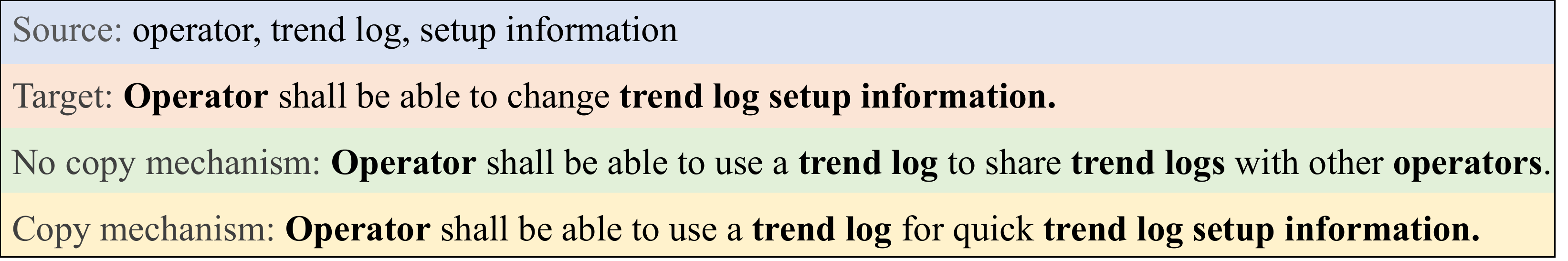} 
\caption{An example with or without copy mechanism}\label{fig:eg4copy}
\end{figure}

We also show the influence of the copy mechanism with an example in Fig. \ref{fig:eg4copy}, and we can see that all keywords, including the multi-term phrases, can occur in the final sentence. Without this mechanism, some keywords are abandoned, in favor of the \emph{more important injected knowledge} and the syntax learned during the pre-training process of UniLM.

Finally, we added syntax-constrained decoding into the settings, and the results are listed in the fifth row of each case in Table \ref{tab:ablationexperiment}. We can see consistent enhancements of the ROUGE-2 results for both cases and a positive effect on the BLEU2 of the BAS domain, and no change to the BLEU2 result of the UAV domain. This indicates that \emph{our syntax decoding is helpful for valid 2-gram generation}, the semantic units with a larger granularity than single-term words.

\section{Discussion}
\label{sec:discussion}

\subsubsection{Threats to validity}
\label{subsec:threats}
\textbf{Conclusion validity} mainly concerns the evaluation metrics. We used BLEU1, BLEU2, and the precision, recall, and F-measure of ROUGE-1, ROUGE-2, and ROUGE-L, which are the most common metrics of NLG \cite{DBLP:conf/emnlp/LinZSZBCR20}\cite{DBLP:journals/corr/abs-2012-10813}\cite{DBLP:conf/emnlp/Sha20}. Moreover, to consider the special syntax constraints of well-formed software requirements \cite{5328509}\cite{8559686}\cite{DBLP:conf/re/GuoZL21}, we designed the RS4RE metric, which measures the syntax distance between the automated generated and target requirements. We also recorded the training and testing time of the automated approaches for practical consideration.

\textbf{Internal validity} concerns the validity of the causal relations between the results and our approaches. We reused the open source code of UniLM, the backbone of our \emph{ReqGen} and the six baselines to ensure the correctness of the implementation. Another threat comes from the measure of RS4RE, which involves the manual annotation of the semantic elements of M-FRDL \cite{DBLP:conf/re/GuoZL21} for the case requirements. In the two sets of requirements, the UAV requirements were annotated by the work of \cite{DBLP:conf/re/GuoZL21}, and we used their original annotations directly. For the BAS domain, our authors performed pair annotations which were then carefully checked individually by the first author of \cite{DBLP:conf/re/GuoZL21}. We believe the annotations are trustworthy.

\textbf{External validity} concerns the generalization of the experimental results to other cases. In this initial study, we experimented with two public datasets from different domains with different scales. All of the requirements and ontologies are from different groups, reflecting different writing styles. However, because of the data limitations, we only performed 10- and 5-fold cross-validation on the UAV and BAS cases. Although the results are promising, more diverse data for further experiments will be needed in future. 

\subsubsection{Implications}

Based on our results, the primary implication is that, like traditional large-scale text-based fine-tuning, concept-net knowledge injection is beneficial for pre-trained models performing  downstream tasks. This means that our approach is potentially helpful for other document generation, such as the description of software design. In addition, multiple knowledge injections into the pre-trained model are better than once, and the knowledge injected into higher layers should be more refined than that injected into the lower layers, just as in the human learning process. Our experiments show that injection of all 5-hop knowledge into the first layer, 2-hop knowledge into the second layer, and only the most relevant 1-hop knowledge into the fourth layer achieves best results on both domains in our task (as shown in Table \ref{tab:timesofmultiinjection}).

The second implication suggested by our study is that automated requirement generation requires related domain knowledge but ``more'' does not always mean ``better''. In the BAS case, 10-frequency filtering is better than no filtering. However, the injected knowledge is also critical because the general UniLM has to learn the domain knowledge for the valid requirement generation. Thus, for the small UAV case, no filtering is the best (as shown in Table \ref{tab:frequencyfiltering}).

\subsubsection{Limitations}
\label{subsubsec:limitations}

Temporarily, we only randomly selected a few keywords as the seeds of \emph{ReqGen} for the requirement statement generation and did not restrict their roles (e.g., specifying that the subject is mandatory). We plan to explore the impact of syntax role configurations, that the given keywords belong to, to the specification generation.

Our method cannot replace the elicitation and analysis of users' requirements. In other words, our aim is only to assist engineers to quickly produce the requirements that they already roughly know (i.e., the keywords). Meanwhile, we cannot guarantee that all the specifications are generated with professional domain words and  clear formal expressions. 

We did not perform an empirical study of the practical usefulness of \emph{ReqGen} in real-world practice. In this initial study, we only evaluated its effectiveness by comparing it with six popular NLG approaches on the common indicators of BLEU, ROUGE and our proposed RS4RE. 

\section{Related work}
\label{sec:relatedWork}

We first review requirements generation research. Then we describe the related work about common NLG.

\subsubsection{Automated requirements generation}
\label{subsubsec:automatedReqGen}

The existing studies on automated requirements generation mainly focus on transforming the (semi-)structured models (e.g., business process model in \cite{turetken2004automating}\cite{DBLP:journals/infsof/CoxPBV05}, i* framework in \cite{DBLP:journals/re/MaidenMJG05}\cite{DBLP:conf/coopis/YuBDM95}, KAOS and Ojectiver in \cite{DBLP:conf/sigsoft/LetierL02}\cite{DBLP:journals/re/LandtsheerLL04}\cite{van2004goal}, UML models in \cite{DBLP:journals/tse/LamsweerdeW98}\cite{DBLP:journals/re/MezianeAA08}\cite{DBLP:conf/re/Berenbach03a}) or other representation (e.g., security goals in \cite{10.1007/s00766-017-0279-5}) into specific syntactic pattern-oriented natural language requirements specifications, based on a set of pre-defined rules. They usually require precise representation of the critical elements, such as the roles, inputs, outputs and their relationships in the business process model \cite{turetken2004automating}, or the security goals expressed as clauses with a main verb + several security criteria + several target assets \cite{10.1007/s00766-017-0279-5}. Both the (semi-) structured inputs and the pre-defined rules restrict the application scope of these approaches. 


Mohamed et al. \cite{9604619} proposed to generate \emph{non-quality} requirements based on grammatical rules and a supplied dictionary, to resolve  small-scale requirements sets for current requirements quality checking work.



There are very few studies on requirements specification generation from a few simple keywords. Most available research on software requirements capturing and generation concerns automatically collecting and/or analyzing requirements-related information with the purpose of \emph{assisting} requirements analysts to \emph{manually} acquire, interpret, and specify the final requirement statements. This information can be requirement-relevant sentences obtained from domain documents \cite{DBLP:conf/re/LianRCZFS16}\cite{ DBLP:conf/re/LianCZ17}\cite{DBLP:journals/infsof/LianLZ20}\cite{LI2015582}\cite{DBLP:conf/re/SleimiCSSBD19}\cite{DBLP:conf/re/SainaniAJG20}, features and their relationships mined from developer online chats\cite{9283914}, online reviews \cite{DBLP:conf/re/AstegherBPS21}\cite{DBLP:conf/re/JohannSBM17} and product descriptions \cite{DBLP:conf/icse/DumitruGHCMCM11}\cite{DBLP:conf/splc/Sree-KumarPC18}, or the classifications on types of user statements \cite{DBLP:conf/re/KhanXLW19}\cite{DBLP:conf/re/Tizard19} and on obligation or non-obligation statements in the contracts\cite{DBLP:conf/re/SainaniAJG20}. 
Besides, better product descriptions can be automatically generated with the techniques of summarization from user reviews\cite{DBLP:conf/www/NovgorodovEGR19}\cite{DBLP:journals/toit/NovgorodovGER20} and website information\cite{DBLP:conf/cikm/EladGNKR19}. 
Although these sources of information are definitely essential for requirements specification, analysts still need to invest significant time and effort interpreting this information and specifying the requirements clearly under the premise of knowing the basic requirements syntax and the domain phrases involved in the requirements.  



\subsubsection{Automated NLG with lexical constraints}
\label{subsec:NLG}

In an early work on lexical constraints generation, Mou et al.\cite{mou2015backward} proposed the backward and forward language model(B/F-LM), and used recurrent neural network to generate previous and subsequent words conditioned on the given word. Liu et al.\cite{DBLP:journals/taslp/LiuFQL19} extended the B/F-LM by introducing a discriminator, but these two methods can generate sentences with only one lexical constraint. To exceed this limitation, Hokamp et al. \cite{DBLP:conf/acl/HokampL17} incorporated constraints by performing a grid beam search in the sentence space. The CGMH framework\cite{DBLP:conf/aaai/MiaoZMYL19} models local transitions (e.g., deletion and insertion) to achieve better fluency, but it is slow to converge. Sha et al. \cite{DBLP:conf/emnlp/Sha20} proposed an unsupervised lexical constraint generation method, in which a series of differentiable loss functions are used to calculate the fluency of the generated sentences and to determine whether they satisfy the constraints. 
Ding et al.\cite{DBLP:conf/emnlp/DingP16} proposed a framework to customize the message content for appealing different individuals.

Recently, the fine-tuning of pre-trained language models has provided more research opportunities for small datasets in many domains.  BERT\cite{devlin2018bert} and RoBERTa\cite{liu2019roberta} use masked language modeling pre-training objectives for deep bidirectional representations by jointly conditioning on both the left and right contexts in all layers. GPT-2/3\cite{radford2019language}\cite{brown2020language} and CTRL\cite{DBLP:journals/corr/abs-1909-05858} are causal language models, which use auto-regressive language to model the target. UniLM\cite{DBLP:conf/nips/00040WWLWGZH19} and GLM\cite{DBLP:conf/acl/DuQLDQY022} combine the advantages of the first two models and use a special mask mechanism so that the tokens in the input sequence can focus on each other, whereas the tokens in the output sequence can focus only on the tokens to the left. MASS\cite{DBLP:conf/icml/SongTQLL19}, T5\cite{DBLP:journals/jmlr/RaffelSRLNMZLL20}, and BART\cite{DBLP:journals/corr/abs-1910-13461} are encoder-decoder models and adopt the standard Transformer structure \cite{DBLP:conf/nips/VaswaniSPUJGKP17}.

\section{Conclusion}
\label{sec:conclusion}

This study aimed to automatically generate requirement statements based on pre-defined keywords. We proposed an approach called \emph{ReqGen} that fine-tunes UniLM by injecting keyword-related knowledge with a repeated emphasis on the most relevant ones, integrates a copy mechanism to ensure the hard constraint of keyword inclusion, and uses syntax-constrained decoding to cater to syntax requirements. Compared with six popular baselines, we showed that \emph{ReqGen} obtains superior performance on the requirements specification generation task.

\bibliographystyle{IEEEtran}
\bibliography{manuscript}

\end{document}